\documentclass{PoS}

\usepackage{amssymb,fontenc,times,mathptmx,epsfig}
\usepackage{psfig,epsf,latexsym,graphicx}

\PoS{PoS(LAT2005)227}

\newcommand{\MS}{\overline{\rm MS}}
\newcommand{\RS}{\rm RS}

\newcommand{\OS}{\rm OS}

\newcommand{\be}{\begin{equation}}
\newcommand{\ee}{\end{equation}}
\newcommand{\bea}{\begin{eqnarray}}
\newcommand{\eea}{\end{eqnarray}}

\newcommand{\nn}{\nonumber}

\newcommand{\als}{\alpha_s}

\sffamily

\def\gev{\ifmmode \mathop{\rm GeV}\nolimits\else {\rm GeV}\fi}
\def\mev{\ifmmode \mathop{\rm MeV}\nolimits\else {\rm MeV}\fi}
\def\kev{\ifmmode \mathop{\rm keV}\nolimits\else {\rm keV}\fi}
\def\ev{\ifmmode \mathop{\rm eV}\nolimits\else {\rm eV}\fi}
\def\siml{{\ \lower-1.2pt\vbox{\hbox{\rlap{$<$}\lower6pt\vbox{\hbox{$\sim$}}}}\ }} 

\title{Lattice and renormalons in heavy quark physics }

\ShortTitle{Lattice and renormalons in heavy quark physics }

\author{\speaker{Antonio Pineda}\thanks{This work is supported 
by MCyT and Feder, FPA 2004-04582-C02-01, and by CIRIT, 2001SGR-00065.}\\
Dept. d'Estructura i Constituents de la Mat\`eria,
\\ U. Barcelona,  
Diagonal 647, E-08028 Barcelona, Catalonia, Spain\\
        E-mail: \email{pineda@ecm.ub.es}}

\abstract{Perturbative expansions of QCD observables in powers of $\alpha_s$
are believed to be asymptotic and non-Borel summable due to the existence of
singularities in the Borel plane (renormalons). This fact is connected with
the factorization of scales (which is inherent to QCD and asymptotic freedom)
and jeopardizes the convergence of the perturbative expansion and the accurate
determination of power-suppressed corrections. This problem is more acute for
physical systems composed by one or more heavy quarks. In lattice regulations,
it reflects on the appearance of power-like divergences in the inverse of the
lattice spacing for a series of quantities ($\bar \Lambda$, gluelump masses,
the singlet and hybrid potentials, ...) making that the continuum limit can
not be reached for them. Nevertheless, all these problems are solved within
the framework of effective field theories with renormalon substraction.
This allows us to obtain convergent perturbative series and to unambiguously
define power corrections. In particular, one can connect with lattice results.
Remarkably enough the dependence on the lattice spacing can be predicted by
perturbation theory. This framework has been applied
 to the prediction of the gluelump masses and the singlet and octet (hybrid)
potentials at short distances, as well as to their comparison with lattice
simulations. Overall, very good agreement with data is obtained.
}

\FullConference{XXIIIrd International Symposium on Lattice Field Theory\\
         25-30 July 2005\\
         Trinity College, Dublin, Ireland}

\begin{document}

\section{Introduction}

\begin{figure}[h!!]
\makebox[1.0cm]{\phantom b}
\put(-30,10){\epsfxsize=6.9truecm \epsfbox{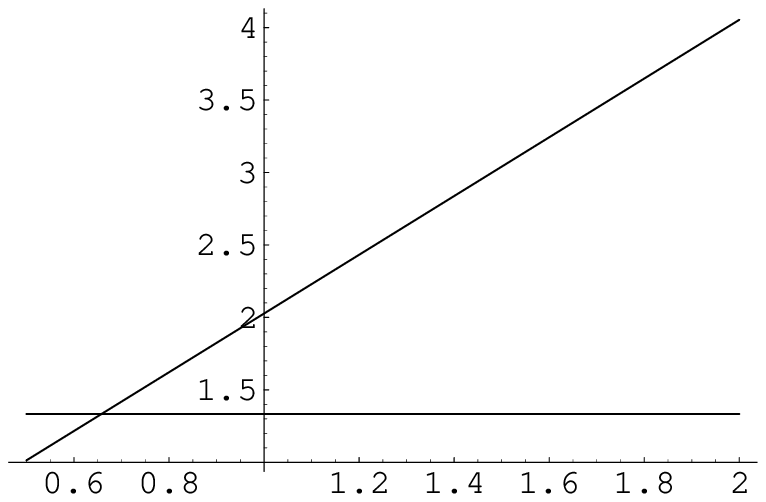}}
\put(210,10){\epsfxsize=6.9truecm \epsfbox{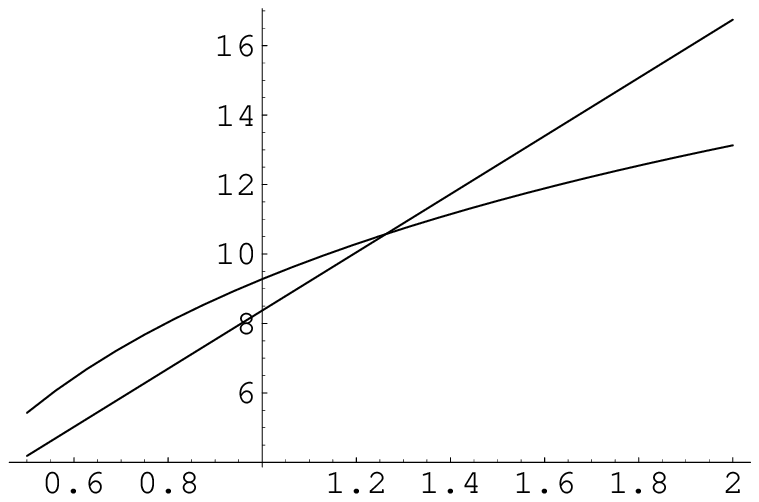}}
\put(-30,-180){\epsfxsize=6.9truecm \epsfbox{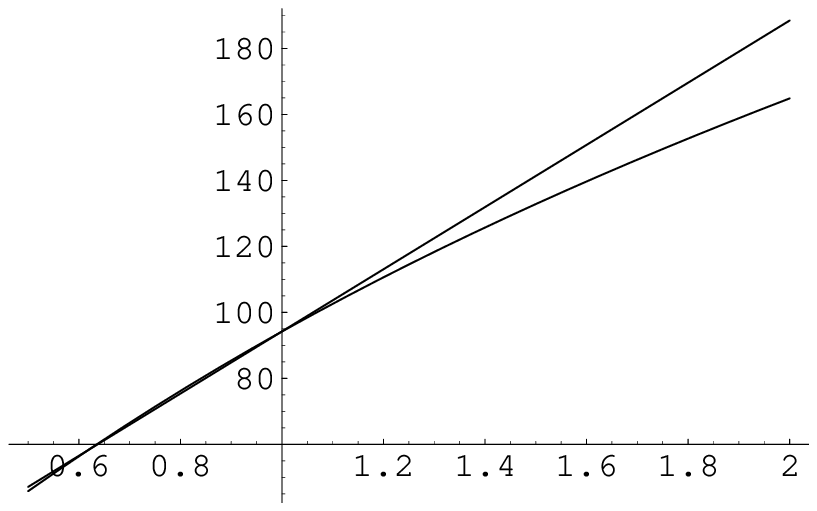}}
\put(210,-180){\epsfxsize=6.9truecm \epsfbox{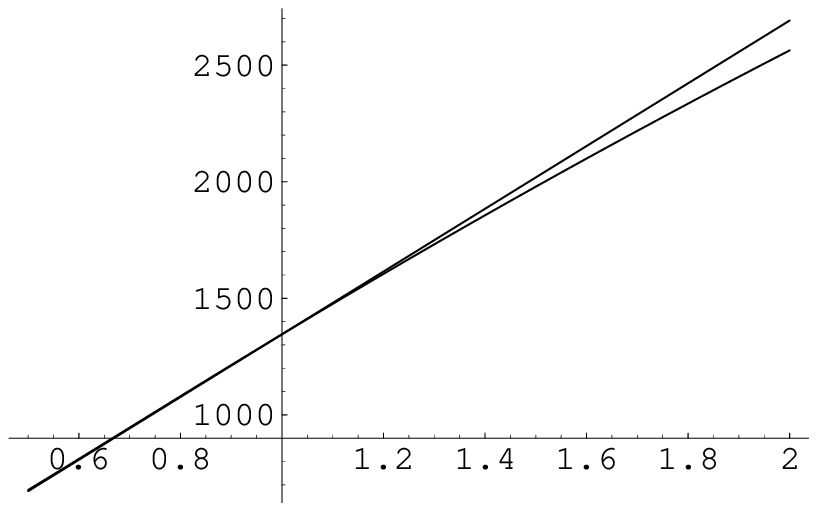}}
\put(140,135){$r_0^{as}$}
\put(140,40){$r_0^{ex}$}
\put(150,1){$\nu/m_{\MS}$}
\put(380,135){$r_1^{as}$}
\put(380,80){$r_1^{ex}$}
\put(390,1){$\nu/m_{\MS}$}
\put(-15,1){$r_0$}
\put(240,1){$r_1$}
\put(-15,-190){$r_2$}
\put(240,-190){$r_3$}
\put(140,-60){$r_2^{as}$}
\put(140,-105){$r_2^{ex}$}
\put(150,-190){$\nu/m_{\MS}$}
\put(380,-60){$r_3^{as}$}
\put(380,-90){$r_3^{ex}$}
\put(390,-190){$\nu/m_{\MS}$}
\caption{Plots of the exact ($r_n^{ex}$) and asymptotic ($r_n^{as}$) 
value of $r_n(\nu)$ at different orders in perturbation theory as a function 
of $\nu/m_{\MS}$. The scale dependence of $r_3^{ex}$ is known exactly. The 
constant term has been fixed using renormalon dominance. }
\label{combinedOS} \vspace{1mm}
\end{figure}

The perturbative series relating the pole and the $\MS$ mass: 
\be
\label{series}
m_{\OS} = m_{\MS} + \sum_{n=0}^\infty r_n(\nu) \als^{n+1}(\nu)\,, 
\ee
suffers from renormalon ambiguities, which makes this series asymptotic 
and non-Borel summable. The behavior of the perturbative expansion at large
orders is dictated by the closest singularity to the origin of its
Borel transform:
\be
B[m_{\OS}](t(u))=N_m\nu {1 \over
(1-2u)^{1+b}}\left(1+c_1(1-2u)
c_2(1-2u)^2+\cdots \right)+({\rm
analytic\; term}),
\ee
where  ($u={\beta_0 t \over 4 \pi}$)
\be\label{borel}
m_{\OS} = m_{\MS} + \int\limits_0^\infty\mbox{d} t \,e^{-t/\als}
\,B[m_{\OS}](t)
\,,
\qquad B[m_{\OS}](t)\equiv \sum_{n=0}^\infty 
r_n \frac{t^n}{n!} . 
\ee
The asymptotic behavior of the coefficients of the perturbative series then reads 
\be
r_n^{as} \stackrel{n\rightarrow\infty}{=} N_m\,\nu\,\left({\beta_0 \over 2\pi}\right)^n
\,{\Gamma(n+1+b) \over
\Gamma(1+b)}
\left(
1+\frac{b}{(n+b)}c_1+\frac{b(b-1)}{(n+b)(n+b-1)}c_2+ \cdots
\right).
\ee
Quite remarkable, it is possible to obtain:\\
{\bf a)} the coefficients $b$, $c_1$, $c_2$,.., exactly through the use of the 
renormalization group \cite{Beneke:1994rs,Pineda:2001zq} 
(actually they will depend on the 
coefficients of the beta function: $\beta_0$, $\beta_1$, ...). 
\\
{\bf b)} approximate determinations of the normalization constant, 
$N_m$, \cite{Pineda:2001zq} by defining new functions with 
improved analitical properties in the Borel plane  \cite{Lee:1996yk}, 
such that for those it 
is possible to perform an analytic expansion in the Borel parameter $u$.
The determination of $N_m$ is quite solid and survives a series of checks, 
see \cite{Pineda:2001zq,Pineda:2002se}:
\begin{itemize}
\item
Good convergence of the perturbative series in $u$ that determines $N_m$.
\item
Mild scale dependence of $N_m$.
\item
Consistency with the determination of $N_{V_s}$, the  
normalization constant of the infrared renormalon of static singlet 
potential $V_s$. $2N_m+N_{V_s} \simeq 0$.
\item
Agreement of the absolute value and scale dependence of the exact and 
asymptotic estimates of the coefficients of the perturbative series. 
See Fig. \ref{combinedOS}.
\end{itemize}

\section{Applications}

As we can see in Fig. \ref{combinedOS}, in heavy quark physics  
the asymptotic behavior sets in at quite low orders in perturbation theory, 
and the powers of $\ln[\nu/m]$ 
effectively exponentiate, becoming a linear power-like 
divergence in the factorization scale\footnote{At this stage the 
similitude with a $1/a$ power-like divergence one would find in lattice 
computations is evident.}:
\be
r_n \stackrel{n\rightarrow\infty}{\sim} m_{\MS}
\left({\beta_0\over2\pi}\right)^n
n!N_m\sum_{s=0}^{n}
{\ln^s[\nu/m_{\MS}] \over s!}
\,.
\ee
The associated lack of convergence of the 
perturbative series at low orders in perturbation theory becomes a problem 
in applications to heavy quark physics. The solution 
proposed in Ref. \cite{Pineda:2001zq} was to shift the $n!$ factorial behavior 
from the perturbative series to the low energy matrix elements, where it 
properly belongs, since this behavior is associated to low energy dynamics. 
For the heavy quark mass, this implies to work with the RS mass:
\be
\displaystyle{m_{\RS}(\nu_f)=m_{\OS}-\delta m_{\RS}=m_{\OS}
-\sum_{n=0}^\infty  N_m\,\nu_f\,\left({\beta_0 \over
2\pi}\right
)^n \als^{n+1}(\nu_f)\,\sum_{k=0}^\infty c_k{\Gamma(n+1+b-k) \over
\Gamma(1+b-k)}}
\,.
\ee
This framework of renormalon subtraction can be applied to any effective 
theory with heavy quarks: HQET, NRQCD, pNRQCD, ... (see \cite{Neubert:1993mb,Brambilla:2004jw} for reviews). 
In this scheme 
some parameters become dependent on the scale, $\nu_f$, and scheme of renormalon 
subtraction. Here we will focus on a series of (quasi-) observables 
that can be studied in the static limit: 
\bea
\langle M_{B/D} \rangle 
&=&
m_{b/c,\RS}(\nu_f)+\bar \Lambda_{\RS}(\nu_f)+{\cal O}(1/m_{b/c,\RS})
\,,
\\
E_s(r)&=&2m_{\RS}(\nu_f)+V_{s,\RS}(r; \nu_f)+{\cal O}(r^2)
\qquad\qquad\qquad\;\; V_s=-C_F\als/r+\cdots\,,
\\
E_H(r)&=&2m_{\RS}(\nu_f)+V_{o,\RS}(r;\nu_f)+\Lambda^{\RS}_H(\nu_f)+{\cal O}(r^2)
\qquad V_o=1/(2N_c)\als/r+\cdots
\,,
\eea
where $H$ labels the hybrid/gluelump state at short distances, and the label 
RS for $V_{s/o}$, $\bar \Lambda$, $\Lambda_H$ means that the leading infrared renormalon 
has been subtracted (added) from the perturbative series. 
The use of this scheme significantly improves the convergence of the 
perturbative series and the agreement with experiment and 
lattice simulations, when available, see 
\cite{Pineda:2001zq,Pineda:2002se,Bali:2003jq}. 
Examples include 
the determination of the 1S bottomonium mass and the agreement with the 
singlet and hybrid potentials computed in the lattice at short distances. 
Another nice example is the accurate description of the dependence on the 
lattice spacing of the static singlet and hybrid potential, $\bar \Lambda$ 
and $\Lambda_H$ using perturbation theory. This is possible because 
one may also consider to do the renormalon subtraction in a scheme 
different from the one presented here (as far as the subtracted quantity 
has the same non-analytic behavior in the Borel plane this would be legitimate). 
Here we would like to connect with the lattice scheme. In practice, this 
means to perform the replacements: 
$\{\RS \rightarrow L, \nu_f \rightarrow 1/a\}$. 
The relation between both schemes is renormalon free and governed by a 
convergent series as we shall next see.
\begin{figure}
\begin{center}
\includegraphics[width=0.69\columnwidth]{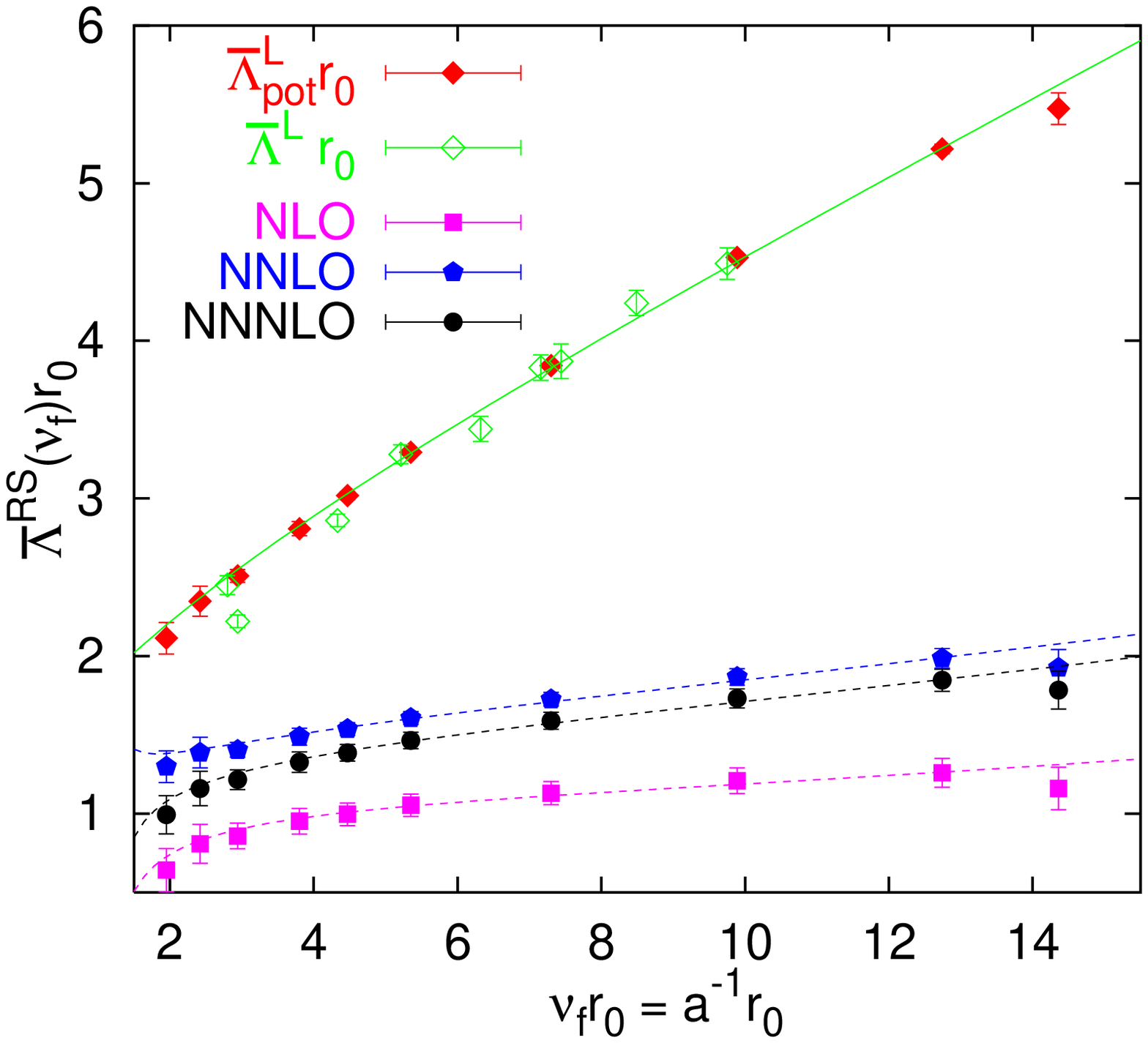}
\includegraphics[width=0.69\columnwidth]{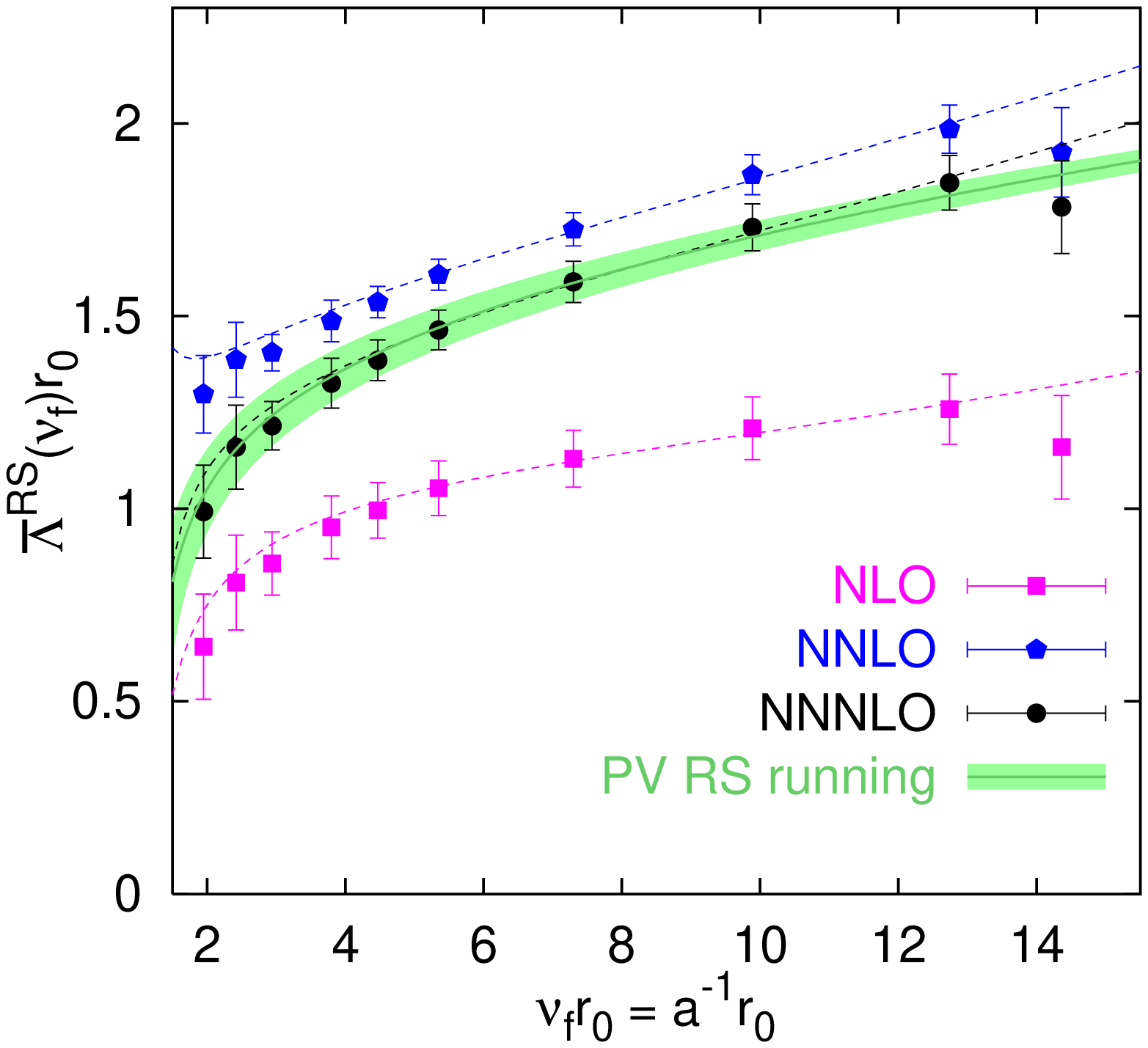}
\end{center}
\caption{\label{potrun}
The first figure shows the binding energy
$\overline{\Lambda}^L_{\rm pot}$ 
(full diamonds \cite{statpot}), in comparison with 
$\overline{\Lambda}^L$ (open diamonds \cite{Duncan:1994uq,barLambda}). 
NLO, NNLO and NNNLO refer to transformations of
$\overline{\Lambda}^L_{\rm pot}$ into the RS scheme
to different orders in perturbation theory. The solid
line corresponds to the NNNLO evaluation in the lattice scheme 
with the central value
$\overline{\Lambda}^{\RS}(\nu_f=9.76\,r_0^{-1})=
1.70\,r_0^{-1}$. The second figures shows 
$\overline{\Lambda}^L_{\rm pot}$ translated into
the RS scheme at NLO (squares), NNLO (pentagons) and NNNLO (circles). 
The solid line corresponds to the principal value running in the RS scheme. 
The error band corresponds to the prediction
$\overline{\Lambda}^{\RS}(9.76\,r_0^{-1})=(1.70\pm 0.04)\,r_0^{-1}$,
and includes the uncertainty due to
$\Lambda_{\MS}=(0.602\pm 0.048)\,r_0^{-1}$. 
For details see \cite{Bali:2003jq}. }
\end{figure}

\section{Perturbative running of $1/a$ and scheme dependence}

Due to the existence of renormalon ambiguities ($1/a$ 
divergences in the lattice language), it is not possible to get 
the continuum limit of the static singlet and octet potential, 
gluelump masses, $\Lambda_H$ and $\bar \Lambda$. Nevertheless, their 
dependence on $1/a$ is free of renormalons and can be 
predicted by perturbation theory:  
\be
\label{runningpot}
2({\bar \Lambda}_L(1/a)-{\bar \Lambda}_L(1/a'))=V_{s,L}(r;1/a)-V_{s,L}(r;1/a')
=C_F\left({1\over a}-{1\over a'}\right)v_1\als+\cdots
\,.
\ee
The coefficients $v_1$, $v_2$, $v_3$ are known in pure gauge theory 
with Wilson action \cite{Duncan:1994uq,latpert}. A similar 
renormalon-free equality can be constructed for the octet potential and 
gluelump masses. 
\bea
\nn
\Lambda_H^L(1/a)- \Lambda_H^L(1/a')
&=&
\left[V_{o,L}(r;1/a)-V_{s,L}(r;1/a)\right]
-
\left[V_{o,L}(r;1/a')-V_{s,L}(r;1/a')\right]
\\
&=&{C_A \over 2}\left({1\over a}-{1\over a'}\right)v'_1\als+\cdots
\,,
\eea
where $v'_1$, $v'_2$ are exactly known and $v'_3$ in the large $N_c$ 
limit. 

We show the plot corresponding to Eq. (\ref{runningpot}) in Fig. 
\ref{potrun}, where the static potential data has been normalized 
to agree with $\bar \Lambda$ at one specific value of the lattice 
spacing 
\begin{equation}
\label{eql1}
\overline{\Lambda}^L_{\rm pot}(a)=\frac{1}{2}V_s^L(r_0;a)+\Delta.
\end{equation}
We can see how nicely the perturbative prediction in the 
lattice scheme (continuous green line in the first plot in Fig. 2) 
agrees with the lattice data \cite{statpot,Duncan:1994uq,barLambda}. 
Significantly, only for one collaboration the slope of the lattice data points 
slightly differs from the prediction of perturbation theory \cite{Duncan:1994uq}. 
We can also see the convergence of the perturbative series relating the 
lattice and the RS scheme, since it is also renormalon free. The final 
value for $\bar \Lambda_{\RS}$, $\bar \Lambda_{\RS}
(\nu_f=2.5r_0^{-1})=1.17r_0^{-1}$, agrees within errors with the result 
obtained directly from experiment using a combined analysis 
of the $\Upsilon(1S)$ and the $B$ meson mass \cite{Pineda:2001zq,Bali:2003jq}, 
$\bar \Lambda_{\RS}(\nu_f=2.5r_0^{-1})=0.92r_0^{-1}$. 

A similar analysis can be performed for the octet potential and gluelump 
masses \cite{Bali:2003jq}. In this case there is less statistics but the 
slope predicted by perturbation theory nicely agrees with the one obtained 
from the gluelump mass $\Lambda_B^L$ measured at different 
lattice spacings \cite{FM}. Moreover, the perturbative series governing the 
change to the RS scheme converges well. Finally, it is quite comforting 
that the same value, within errors, is obtained for $\Lambda_B$ from either 
the hybrid potential, $\Lambda_B^{\RS}(\nu_f=2.5r_0^{-1})=2.25r_0^{-1}$, 
or from the direct lattice determination, $\Lambda_B^{\RS}(\nu_f=2.5r_0^{-1})=2.31r_0^{-1}$.
All these findings can be found summarized in Fig. 14 in \cite{Bali:2003jq}.

In brief, we can relate processes computed with different scales/schemes 
using well-behaved (renormalon free) perurbative series:
\begin{eqnarray*}
\,\left.
\begin{array}{ll}
&
\displaystyle{\bar \Lambda_L(1/a) \qquad \Longleftrightarrow \qquad 
{\bar \Lambda}_L(1/a')}
\\
&
\displaystyle{\;\Updownarrow
(\nu_f=1/a)
\qquad
\qquad
\Updownarrow(\nu_f'=1/a')}
\\
&
\displaystyle{\bar \Lambda_{\RS}(\nu_f) \qquad \Longleftrightarrow \qquad 
{\bar \Lambda}_{\RS}(\nu'_f)}
\end{array} \right\} 
{\rm The\; circle\; can\; be\; closed\; using\; perturbation\; theory}
\end{eqnarray*}
and a similar circle applies for $V_s^L(r;1/a)$, $V_o^L(r;1/a)$ and  
$\Lambda_H^L(1/a)$.

\section{Conclusions}

We have accumulated a lot of evidence in favour of the renormalon 
dominance in heavy quark physics. A proper handle of these effects 
appears to be crucial to accurately describe either lattice or experiment. 
We point out that the dependence on the lattice spacing can be obtained 
from perturbation theory with good accuracy and well controlled 
errors. Therefore, lattice simulations in heavy quark physics can be 
peformed with quite coarse lattices and yet obtain accurate results.
This may have important consequences to diminish errors in lattice 
simulations.

\end{document}